\documentstyle[
pre,
twocolumn,
aps]{revtex}
\begin{document}
\title{Collective Behavior of Asperities in Dry Friction at 
	Small Velocities}
\author{Franti\v{s}ek Slanina}
\address{Institute of Physics,
	Academy of Sciences of the Czech Republic
	Na~Slovance~2, CZ-18040~Praha,
	Czech Republic\\
	and Center for Theoretical Study
	Jilsk\'a~1,CZ-11000~Praha, Czech
	Republic\\
	e-mail: slanina@fzu.cz}
\date{\today}
\maketitle

\begin{abstract} 
We investigate a simple model of dry friction based on extremal
dynamics of asperities. At small  velocities, correlations
develop between the asperities, whose range becomes infinite in the
limit of infinitely slow driving, where the system is self-organized
critical. This collective phenomenon leads to 
effective aging of the asperities and results in velocity dependence of
the friction force in the form $F\sim 1- \exp(-1/v)$.
\end{abstract}

PACS number(s): 46.30.Pa; 64.60.Lx

\section{Introduction}

Phenomena connected with mechanical properties of complex systems are
subject of intensive study in the last decade. Generally speaking, the
difficulty stems from the fact that both macroscopic scale and
mesoscopic scale are important. For example, the contact area of two
grains of sand is a mesoscopic object, but its properties result in
macroscopic behavior of a sand heap. Among the whole family of such
problems, the dry 
friction emerged in recent years as a hot subject. Besides the  
intrinsic interest in dynamics of contact interfaces sliding on top of
each other, there are various systems studied recently, in which
friction forces are dominant interactions determining the behavior.
As examples, we may note two notoriously known phenomena: sand heaps
and earthquakes. Equilibrium stress
distribution in heaps of granular materials exhibits complicated
localized structures \cite{ja_na_be_96,ra_je_mo_ro_96}. 
Dynamics of tectonic plates gives rise to power-law distribution of
earthquakes, formulated in Gutenberg-Richter law
\cite{ba_ta_89,ola_fe_chri_92}. A one-dimensional counterpart of
friction is {\it e. g. } the dislocation movement, 
which is responsible for the plasticity of metals. 

At least three regimes of friction may be distinguished. First, dry
friction corresponds to tangential force acting on the contact of two
macroscopic solid bodies. The slot between the bodies is empty. The
friction emerges as a result of the rheological properties of the
sliding bodies both at the macroscopic and mesoscopic scale.
Second case, the lubricated friction, differs in the fact, that the
slot between bodies is filled with a liquid and the mechanical
properties of mesoscopic portions of the lubricant are responsible for
friction. Third, friction of a single microscopic tip on a surface
may be measured, which explores the microscopic properties of the
surface \cite{sch_zw_ko_wi_97}. Here we concentrate on the first
possibility, dry friction.

Dry friction is intensively studied today, both experimentally and
theoretically  
\cite{he_bau_pe_ca_ca_94,per_to_96}. 
The commonly accepted picture of friction is based on the
dynamics of 
a system of mesoscopic contacts, called asperities, scattered on the
surface of the sliding bodies
\cite{ca_no_96,ca_ve_97a,vo_na_97,ta_ro_97,ta_gou_ro_98}. The typical
size of asperities is constant, while their number is proportional to
normal load. Hence the Amontons-Coulomb law, stating that the apparent
contact area of two macroscopic bodies does not matter.

However, many features are not well understood,
{\it e. g. } the velocity dependence of the friction force. It is
explained either as a consequence of the plasticity of the asperities,
which is considered as a thermally activated process
\cite{he_bau_pe_ca_ca_94}  (this phenomenon
is called aging of the asperities) or purely geometrically, based on
self-affine shape of the surfaces \cite{vo_na_97}. Within the approach
based on 
plasticity, logarithmic dependence of the age of the asperity on time
is supposed on the basis of experimental data, which suggest
logarithmic velocity dependence of the friction force. On the other hand,
geometrical approach gives friction force proportional to $v^{-1}$ for
large velocities, while the behavior for small velocities depends on
the fractal geometry of the surface.

In the description of the process of friction two levels may be
distinguished. On the global level, the 
averaged effect of asperities can be successfully described using the
elasto-plastic model developed by Caroli, Nozi\'eres and Velick\'y
\cite{ca_no_96,ca_ve_97a}. This approach is effectively a single-site one.
Only one asperity is changing its state and
effect of all other asperities is described by effective surrounding
medium. The 
scheme resembles remotely the coherent potential approximation (CPA) 
for electronic structure of alloys.
The spatio-temporal correlations are considered to be of very short
range, and the mutually sliding surfaces behave in uniform way.

On the other hand, the local, short time level of description must
take into account processes which happen at several (or many)
asperities simultaneously, or within a very short period of time, so
that they cannot be 
considered as uncorrelated. Several approaches in this direction were
already proposed, based on geometrical considerations
\cite{vo_na_97,ta_ro_97}, on Frenkel-Kontorova \cite{we_el_97},
Burridge-Knopoff and 
train models \cite{elmer_96,elmer_97} or on an extremal dynamics model
with elastic interactions \cite{ta_gou_ro_98}.

The extremal dynamics (ED) models are very appealing, because they may
grasp the ``skeleton'' of the problem, despite their simplicity and
rudimentary nature. Generally, ED is based on the
assumption that only one site is evolving during one time step, namely
the site which has maximum (or minimum: it depends on the model in
question) of the dynamical variable determining the state of the
system. However, the price to pay is that the time scale fixed by the
frequency of the updates of single sites is not directly related to
the real time measured in an experiment.

Extremal dynamics models were successfully used in modeling various
systems, like invasion percolation, biological evolution
\cite{pa_ma_ba_96}, earthquakes \cite{ito_95} or dislocation movement
\cite{za_92,ma_zha_95}. The model we propose here is based on the ideas of
ED models, adapted to the fact, that in friction we are
interested in macroscopic movement with non-zero velocity, while most
of ED models are appropriate to the case of infinitely slow movement.

Shortly,
he evolution of our model
proceeds at the most susceptible asperity, namely the asperity which
bears maximum stress. 
A small
mechanical perturbation, like release of stress at single site,
may result in a burst of activity of large spatio-temporal extent.
Following the terminology used in the theories of self-organized criticality
\cite{pa_ma_ba_96},  we
will call such spatio-temporal areas of activity avalanches. The
correlations present 
in the model will be described through statistical properties of the
avalanches. 

From time to time, the ED of asperities is interrupted by a
macroscopic ``slip'' of the body as a whole, in which all asperities
are completely renewed. By combination of ED evolution with such
macroscopic slips, we introduce non-zero macroscopic sliding velocity
into the model.

The rest of the article is organized as follows. In the section II the
model is defined and the interpretation of the model parameters is
given, in sect. III the presence of self-organized criticality is
investigated for in the case of zero macroscopic velocity, while the
effect of non-zero velocity on the breakdown of SOC as well as the
velocity dependence of the friction force is investigated in
sect. IV. The section V summarizes the results and draws conclusions
from them.

\section{The model}

We propose the following model.
There are $N$ point contacts, asperities, each with stress $b$. 
The quantity $b$ will be interpreted as the elastic energy stored in
the asperity.
The model is
one-dimensional (the generalization to realistic two-dimensional case
is straightforward) with periodic boundary conditions , so the points
form a closed ring. In each step, the point with highest stress $b_{max}$
is found and released. The release of the stress means, that the point
is removed. However, in order to keep the number of points constant,
new point is introduced somewhere in the system. 

As a zeroth
approximation, the location of new point may be
chosen at random.
However, in reality the
position of the new contact is determined by the detailed structure of
the the surfaces of slide and track. The new contact is established at
such place, where the surfaces are closest one to another. So, another
number, $d$ is attributed  
to each point representing the width of the slot between the surfaces,
waiting in the vicinity of the asperity for 
further updates (the {\it actual} slot directly on the asperity is
zero, of course), and in the 
update the location of minimum $d$ is found. 
Here a new asperity is reintroduced. The values of $b$ and $d$ of the
neighbors of the old and new sites are also updated. Generally, each
site has $K-1$ neighbors which 
are affected. For
simplicity, we assume $K=2$, and update only one neighbor (the right one).   

Let us allow for very slow motion of the slider as a whole. The energy
stored in the released asperity may be transferred entirely to other
asperities, or a part of it may be converted into kinetic energy
$E$. It
may also happen that some of the kinetic energy is returned back to
elastic energy of some asperities. 
It is natural to expect, that at higher velocities, the number of
asperities affected by the transfer of the kinetic energy to the
elastic one will be larger. We simplify this dependence by saying, that
for $E<E_{\rm thr}$ only the nearest neighbors are
affected, while for $E\ge E_{\rm thr}$ the slider slips
macroscopically over average distance $x_{\rm slip}$. 
The average duration of the slip is $T_{\rm slip}$ and after that time 
 all parameters $b$ and $d$ of all
asperities are newly attributed at random and $E$ is set to
0. Then, the dynamics starts again. In this process, the kinetic energy
$\simeq E_{\rm thr}$ the system 
had before the slip is dissipated. This makes a difference with the
theories of one asperity dynamics, where the energy is dissipated
immediately after release of a single asperity.
In our model we do not describe the processes which happen during the
slip, {\it e. g. } we do not examine the energy dissipated in course
of the slip. Similarly, we do not calculate the physical velocity
corresponding to the kinetic energy $E$ during the ED evolution. So,
we isolate only those contributions to the friction force and the
macroscopic slider velocity, which originate in the ED process
interrupted by instantaneous slips.

The average macroscopic velocity
$\Delta v$
stemming from the slips depends on the average time
interval between two subsequent slips. We may determine this quantity
$\overline{\Delta t}$ in the time units of the
extremal dynamics process. Its relation to physical time is not
straightforward, but we suppose that this ambiguity affects only
units, in which we measure time and not the general dependence of the
friction force on velocity. Thus, we write simply 
\begin{equation}
  \Delta v =1/\overline{\Delta t}\; 
\label{eq:dv}
\end{equation}
which corresponds to taking the average slip length $x_{\rm slip}$ as
length unit and average time needed to update single asperity as a
time unit.
The contribution $\Delta v$ from the ED process is dominant if the
time between slips is much larger than the duration of the slip,
$\overline{\Delta t} \gg T_{\rm slip}$ ({\it i. e. } slips are
instantaneous events) and the real length trvelled between slips
during the ED dynamics $x_{\rm ED}$ is much shorter than the slip
length, $x_{\rm ED} \ll x_{\rm slip}$.

The contribution $\Delta F_{\rm
fri}$ to the friction force coming from this process is then
proportional to the energy dissipated in one slip. Because we are
using arbitrary units, we identify 
\begin{equation}
  \Delta F_{\rm fri} = E_{\rm thr}\; .
\label{eq:df}
\end{equation}

Let us now describe the extremal dynamics of the model more formally.
The model consists of $N$ sites connected in ring topology. Each site
$i\in \{1,2,...,N\}$ is connected to its right
neighbor $r(i)$. 
The state of the model is described by the set
$(E,b_1,b_2,...,b_N,d_1,d_2,...,d_N)$ and the function
$r(i)$ which describe the connectivity of the sites. 
At the beginning, $E=0$ and
both $b_i$ and $d_i$ are uniformly distributed in the interval (0,1).
The updating steps are the following. 

\vspace{3mm}
(i) Find maximum stress $b_{\rm max}=\max_i(b_{i})$ 
located at site $i_{\rm max}$.
Remember its old right neighbor $i_{\rm
old}=r(i_{\rm max})$. 

(ii) Find minimum slot $d_{\rm min}$ at site $i_{\rm min}$. 

(iii) Change of connectivity: The site $i_{\rm max}$ is removed by
cutting its 
links to left and right nearest neighbors and 
re-inserted between $i_{\rm min}$ and the
site next to it on the ring. It will have new right neighbor $i_{\rm
new} = r(i_{\rm max}) = r(i_{\rm min})$, and then set $r(i_{\rm
min})=i_{\rm max}$.

(iv) Kinetic effects: Set \\
$E^\prime  = E+\delta\cdot b_{\rm max}$,
$b_{\rm max}^\prime = (1-\delta)b_{\rm max}$,\\
$\Delta_1 = (b_{\rm M}-b_{i_{\rm old}})\theta(b_{\rm M}-b_{i_{\rm old}})$,
$\Delta_2 = (b_{\rm M}-b_{i_{\rm new}})\theta(b_{\rm M}-b_{i_{\rm new}})$\\
If $E^\prime > \Delta_1+\Delta_2$, we set\\
$E = E^\prime-\Delta_1-\Delta_2$,
$b_{i_{\rm old}}^\prime = b_{i_{\rm old}}+\Delta_1$,
$b_{i_{\rm new}}^\prime = b_{i_{\rm new}}+\Delta_2$.\\
if not, we set \\
$E = 0$,
$b_{i_{\rm old}}^\prime = b_{i_{\rm old}}+E^\prime/2$,
$b_{i_{\rm new}}^\prime = b_{i_{\rm new}}+E^\prime/2$.

(v) Stress redistribution: For $r_1,r_2$ random numbers distributed
uniformly in the triangle $0<r1<r2<1$  we set \\
$b_{i_{\rm max}} = r_1 b_{\rm max}^\prime$,
$b_{i_{\rm old}} = b_{i_{\rm old}}^\prime + (r_2-r_1) b_{\rm max}^\prime$,
$b_{i_{\rm new}} = b_{i_{\rm new}}^\prime + (1-r_2) b_{\rm max}^\prime$.

(vi) New values of slots $d$ are attributed to old and new neighbors as
well as to site $i_{max}$, taking random numbers uniformly distributed
in the interval (0,1).

(vii) If $E \ge E_{\rm thr}$, slip occurs, which means that
E is set to 0 and $b_i$ and $d_i$ distributed uniformly in the
interval (0,1).

\vspace{3mm}
The kinetic effects and the slip involve several parameters. First,
the parameter $\delta$ describes how much of the elastic energy tends
to be converted into the kinetic energy. This parameter can be
interpreted as a measure of the spatial density of asperities: in the
elementary process of freeing a particular asperity, more elastic
energy may be converted into the kinetic one, if the density of
asperities is lower. If $\delta = 0$, the kinetic effects are turned off.

The parameter $b_{\rm M}$ is the limit, up to which an asperity can
absorb a portion of kinetic energy and convert it back to elastic
energy. It should be the property of the surface itself, without any
resort to the load and velocity of the slider. If $\delta=0$, the
parameter $b_{\rm M}$ does not enter the model.

The slip is determined by the parameter $E_{\rm thr}$. In more
realistic description, it would be necessary to introduce the function
$R(E)$ which would count the number of sites, including the extremal site
$i_{\max}$, which are to be
updated, provided the kinetic energy has 
the value $E$. Here we take the simplest form
$R(E)=3+(N-3)\theta(E-E_{\rm thr})$.
Even this parameter should be the property of the surface,
irrespectively of the load and velocity.

Finally, we comment on the interpretation of the quantity $N$, the
average number of asperities. We suppose that it may serve as a
measure of the external load. While $N$ does not depend on the
apparent contact area of the slider and the track, the asperity
density, which is hidden in the parameter $\delta$, does. 

So, in order to conform with the Amontons-Coulomb law, we expect that
the surface properties will enter the velocity dependence of the
friction force through the parameter $b_{\rm M}$ and combinations
$E_{\rm thr}/N$ and $\delta/N$. We will see later that it is exactly
the case.

\begin{figure}[hb]
  \centering
  \vspace*{80mm}
  \includegraphics{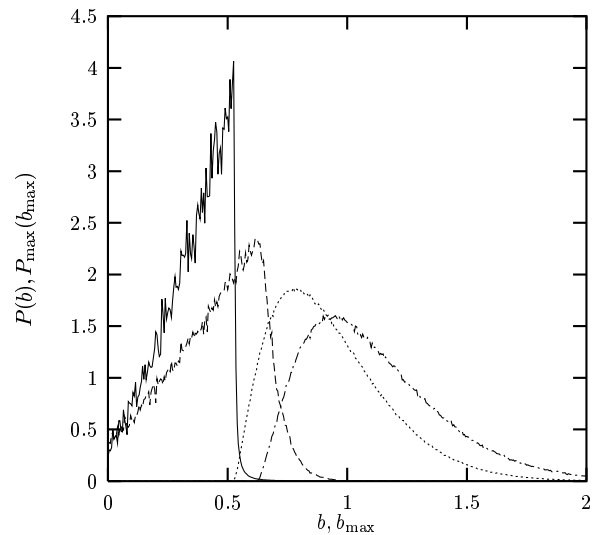}
  \caption{Distribution of stresses $P(b)$ and maximum stresses
$P_{\rm max}(b_{\rm max})$ for $N=1000$, $\delta=0.01$, $b_{\rm M} =
  0.9$. The energy 
threshold is infinite ( full line for $P(b)$ and dotted for $P_{\rm max}(b_{\rm
max})$) and 
$E_{\rm thr} = 0.08$ ( dashed line for $P(b)$ and dash-dotted for
$P_{\rm max}(b_{\rm max})$). Number of steps is $10^6$.}  
  \label{fig:distr}
\end{figure}

\section{Infinitely slow movement regime}

Let us first investigate the case in which no slips are allowed, which
can be expressed by limit value $E_{\rm thr} = \infty$. In this case,
the macroscopic movement is infinitely slow. If the elastic
energy could not be transformed into kinetic energy $E$, {\it i. e. }
if $\delta=0$, the model would be a slightly more complicated version of the
Zaitsev model for dislocation movement \cite{za_92}, which is known
to be self-organized critical. The criticality manifested by power-law
distribution of avalanche sizes is due to infinitely slow driving. It
is natural to expect SOC also in our model for $\delta=0$. However,
even for $\delta > 0$ the condition of infinitely slow driving, which
means technically that only one asperity is updated at a time, is also
satisfied and SOC is expected as well. 

We simulated systems of size $N=1000$. The first quantity we measured
was the probability
distribution of the stresses, $P(b)$ and maximum stresses $P_{\rm
max}(b_{\rm max})$. The function $P(b)$ is continuous up to a critical
value 
$b=b_{\rm c}$ and then suddenly drops to zero, which is behavior
common in SOC extremal dynamics models. The value of $b_{\rm c}$
depends on $\delta$. The typical behavior is shown in Fig.
\ref{fig:distr} for $\delta = 0.01$.

A fingerprint of self-organized criticality is {\it e. g. } the
scaling behavior of the forward $\lambda$-avalanche sizes
\begin{equation}
P_{\rm fwd}(s) = s^{-\tau}g(s|\lambda-\lambda_c|^{1/\sigma})\;\; .
\label{eq:aval-scaling}
\end{equation}
The $\lambda$-avalanche starts when $b_{\rm max}$ exceeds the value
$\lambda$ and ends when  $b_{\rm max}$ drops below the value
$\lambda$ again. The size $s$ of the avalanche is number of update
steps from the start to the end of the avalanche. For numerical
reasons it is simpler to investigate scaling of integrated
distribution,
$P^>_{\rm fwd}(s) = \int_s^\infty {\rm d}\bar{s}\, P_{\rm
fwd}(\bar{s})$ from which the exponents $\tau$ and $\sigma$ can be
determined. 

\begin{figure}[hb]
  \centering
  \vspace*{80mm}
  \includegraphics{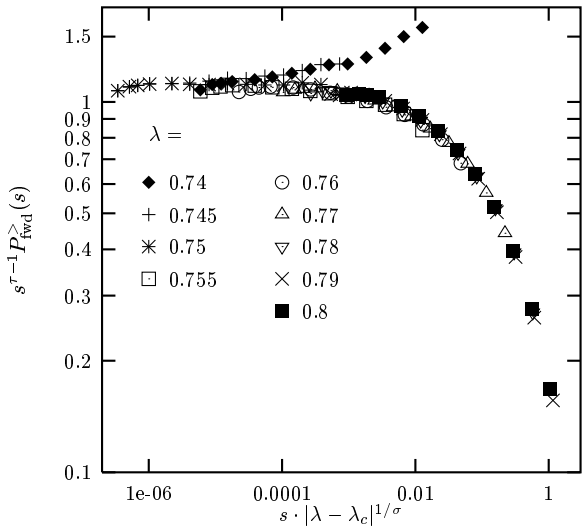}
  \caption{Rescaled forward avalanche distribution for $N=1000$,
$\delta = 0$. The critical threshold is
  $\lambda_c = 0.7475$ and the 
scaling exponents $\tau=1.28$ and $1/\sigma = 2.6$. Number of steps is
$10^8$. The corresponding thresholds $\lambda$ are indicated next to
the symbols in the legend.}
  \label{fig:aval0}
\end{figure}

\begin{figure}[hb]
  \centering
  \vspace*{80mm}
  \includegraphics{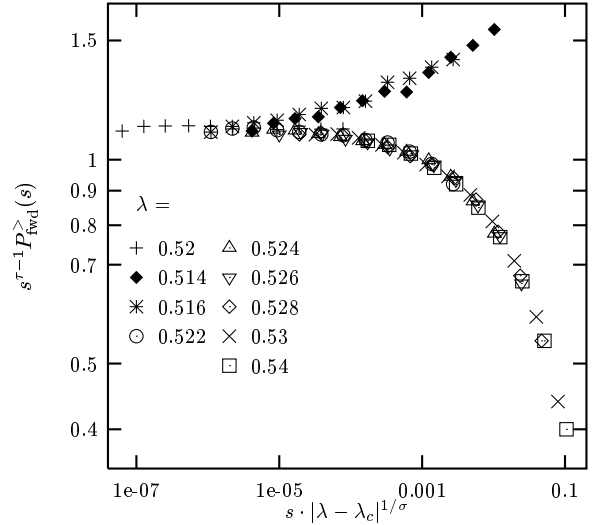}
  \caption{Rescaled forward avalanche distribution for $N=1000$,
        $\delta = 0.001$, $b_{\rm M} =   0.9$. The critical 
        threshold is
        $\lambda_c = 0.519$ and the 
        scaling exponents $\tau=1.27$ and $1/\sigma = 2.6$. 
        Number of steps is
        $10^8$. The corresponding thresholds 
        $\lambda$ are indicated next to
        the symbols in the legend.}
  \label{fig:aval001}
\end{figure}

The Fig. \ref{fig:aval0} and Fig. \ref{fig:aval001} show the data
collapse which confirms the scaling of the form
(\ref{eq:aval-scaling}). Best collapse was obtained for the following
values of the parameters: 
(a) for $\delta=0$ we have $\lambda_c = 0.7475$ and $\tau=1.28$ and
$1/\sigma = 2.6$. 
and (b) for $\delta=0.001$ and $b_{\rm M}= 0.9$ we have $\lambda_c = 0.519$ and $\tau =
1.27$, $1/\sigma = 2.6$.

There is a minor difference in the exponent $\tau$ giving best fit
for $\delta = 0$ and $\delta = 0.001$. However, we believe that this
difference is within numerical uncertainty of the results and the
model belongs to the same universality class irrespectively of
parameter $\delta$.

By qualitative inspection of the quality of the data collapse for
different choices of the exponents, we estimate the error bars. Thus,
we finish with the following critical exponents of our model:
\begin{equation}
\tau = 1.27 \pm 0.02,\;\; \sigma = 0.38\pm 0.02\;\; .
\end{equation}

The forward avalanche exponent $\tau$ is greater than in the 1D
Zaitsev model \cite{za_92,pa_ma_ba_96}, but close to the Sneppen
interface growth model \cite{sneppen_92,pa_ma_ba_96}. Another 1D model
to be compared is the CDW model of Olami \cite{olami_93}, and
anisotropic interface depinning model of ref. \cite{ma_zha_95} which
have however 
significantly larger exponent $\tau$. The closest universality
class seems to be the one of the Sneppen model ($\tau = 1.26$), but
the value of $\sigma = 0.35$ in this class is smaller than in our
model. 

Whether this difference is due to finite-size effect or the two
models are in different universality class cannot be stated with
certainty from our present data. Instead, we would like to stress a
structural similarity of the two models, which may explain the
similarity of exponents. Contrary to usual interface growth models
\cite{ha_zha_95}, the Sneppen model is a non-local one. After a single
growth event (deposition of a single particle), an unbounded sequence
of further steps is performed in order to re-establish the single-step
property of the interface. So, the range of interactions fluctuates
during the evolution, according to the actual configuration of the
interface. Similarly, the Zaitsev model, like most of other extremal
dynamics models is local in the sense that after finding an extremal
site, its neighbors are also updated, while the range of neighborhood
is fixed. On the contrary, our model, like the Sneppen model, does not
have fixed range of interactions, but it is established by the
position of the minimum
of the quantity $d$ (the slot). We simulated also a version, in which
the site, where new asperity is inserted, is chosen at random, instead
of using the slot $d$. In this case we observed mean-field behavior
characterized by exponents $\tau=1.5$, $\sigma=0.5$.

\section{Friction at non-zero velocity}

\begin{figure}[hb]
  \centering
  \vspace*{80mm}
  \includegraphics{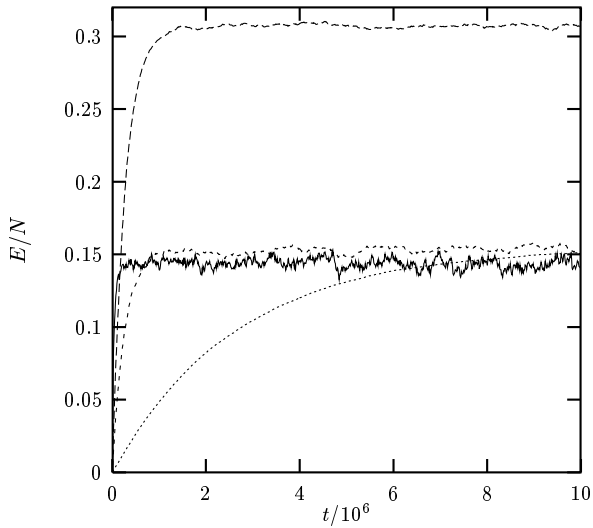}
  \caption{Time evolution of the kinetic energy per asperity. The
           parameters are as follows:
           $N= 1000$, $\delta = 0.01$, 
           $b_{\rm M}= 0.9$ (full line)  
           $N= 10^4$, $\delta = 0.001$, 
           $b_{\rm M}= 0.9$ (dotted line) 
           $N= 1000$, $\delta = 0.001$, 
           $b_{\rm M}= 0.5$ (long dashed line) 
           $N= 1000$, $\delta = 0.001$, 
           $b_{\rm M}= 0.9$ (short   dashed line).}
  \label{fig:ener}
\end{figure}

In the preceding section we dealt with stationary properties of the
model. In order to account for macroscopic movement, transient
properties are of interest. First, we investigate the evolution of the
kinetic energy $E$ and its approach to the stationary value
$E_\infty$, if we forbid the slips, {\it i. e. } $E_{\rm thr} =
\infty$. In Fig. \ref{fig:ener} we
show the time evolution of 
$E/N$ for different values of the model parameters $\delta$, $b_{\rm
M}$ and number of asperities $N$. 

The most important observation is
that the stationary value $E_\infty/N$ depends on $b_{\rm M}$, while
the dependence on $\delta$ and $N$ is within the noise level. (We
observe that both large $N$ and small $\delta$ suppress the relative
fluctuations of the kinetic energy around the stationary value.) The
physical significance is clear: the static friction force, which is
according to eq. (\ref{eq:df}) equal to $E_\infty$, is proportional to
$N$, which is in turn proportional to normal load. Thus, we recover the
Amontons-Coulomb law for static friction.

\begin{figure}[hb]
  \centering
  \vspace*{80mm}
  \includegraphics{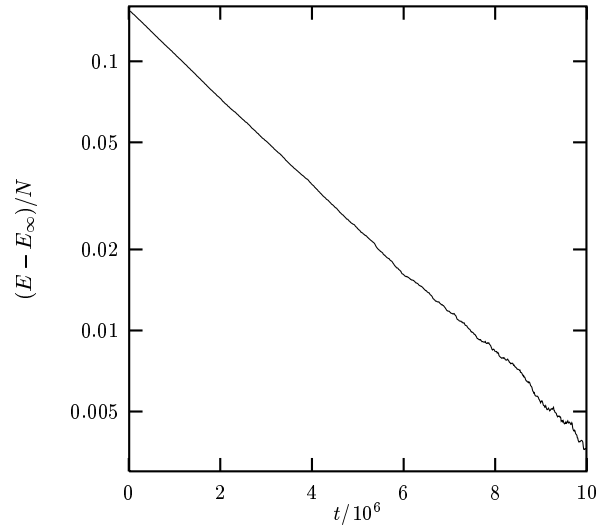}
  \caption{Approach of the kinetic energy to its stationary value, for
        $N=10^4$, $\delta = 0.001$, $b_{\rm M}= 0.9$. Stationary value
        is taken as $E_\infty/N = 0.155$.
          }
  \label{fig:approach}
\end{figure}

The approach of the kinetic energy to its stationary value is
exponential, as is demonstrated in the Fig. \ref{fig:approach}. This
type of approach is directly reflected in the velocity dependence of
the friction force, as we will see below.

\begin{figure}[hb]
  \centering
  \vspace*{80mm}
  \includegraphics{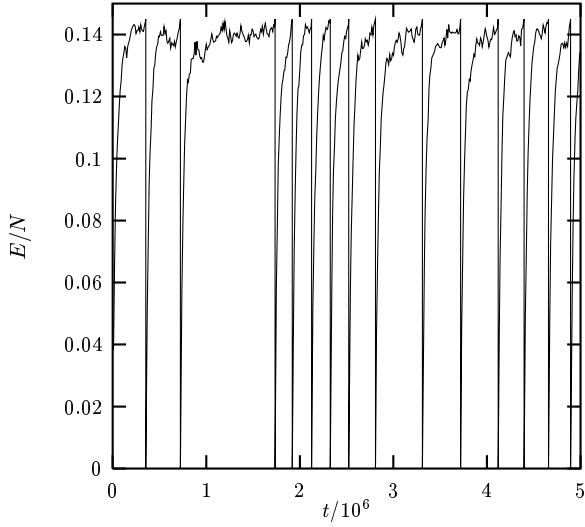}
  \caption{Time dependence of the kinetic energy $E$, for
  $N=10^3$, $\delta = 0.1$ and $b_M=0.9$. The slips occur in the
  moments when $E$ drops to 0.}
  \label{fig:slips}
\end{figure}

\begin{figure}[hb]
  \centering
  \vspace*{80mm}
  \includegraphics{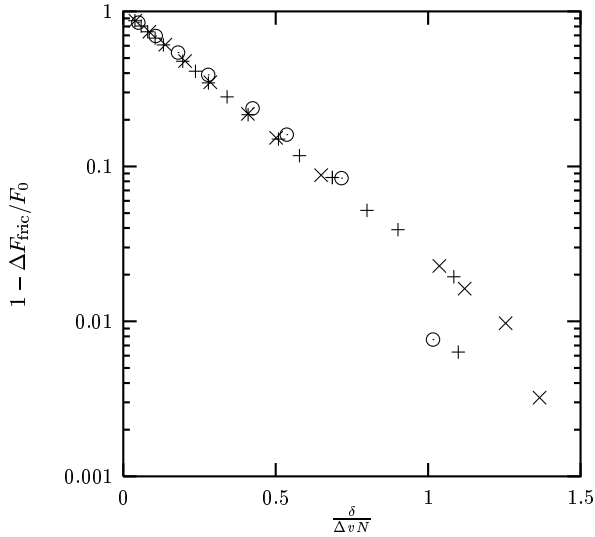}
  \caption{Velocity dependence of the friction force, for $N=10^3$,
  $\delta=0.001$, $b_M=0.9$ ($+$), $\delta=0.001$,
  $b_M=0.5$ ($\times$),$\delta=0.01$, $b_M=0.9$ ($\circ$).}
  \label{fig:force}
\end{figure}

If we set the threshold $E_{\rm thr}<E_\infty$, quasi-periodic
behavior is observed: the kinetic energy grows, until it reaches the
value of the threshold, and then the system is reinitialized. This
regime is illustrated in the Fig. \ref{fig:slips}. If the threshold is
close to $E_\infty$, the slips are less regular, due to fluctuations,
but for smaller vales of the threshold, the slips occur with fixed
frequency. The mean number of steps $\Delta t$ between slips is
determined by the way how $E$ approaches the stationary value. Because
$E_{\rm thr}$ is related to the friction force by (\ref{eq:df}) and mean
period of slips to the velocity, according to (\ref{eq:dv}), the
velocity dependence of the friction force is measurable in our model.
The Fig. \ref{fig:force}
shows the results for various $\delta$ and $b_{\rm M}$.
If we denote $F_0 = E_\infty$ the static friction force, we observe by
plotting the velocity dependence in semi-logarithmic scale that the
following law is well satisfied
\begin{equation}
\Delta F_{\rm fri} = F_0\cdot (1-\exp(- A {\delta \over \Delta v
N}))
\label{eq:dfvsdv}
\end{equation}
with some constant $A$ characteristic of the model. We have found
$A=3.6\pm 0.3$.
The deviations from the above dependence for $\Delta v < \delta/N$ are
due to time fluctuations of $E$, which lead to less regular
slips. However, as we already mentioned, the relative fluctuations
decrease with $N$, so we expect the dependence (\ref{eq:dfvsdv}) to
hold for all velocities in thermodynamic limit $N\to\infty$.

As $\delta$ is interpreted as a quantity proportional to the density
of asperities, the ratio $\delta/N$ does not depend on $N$, {\it
i. e. } on the normal load. Because also $F_0$ was found to be
proportional to $N$, the form of (\ref{eq:dfvsdv}) conforms with
Amontons-Coulomb law for all $\Delta v$.

For large velocities the friction force decreases as $\Delta F_{\rm
fri}\sim 1/\Delta v$. The same velocity dependence was found also using
different approach \cite{vo_na_97}.

Now we turn to the influence of the macroscopic movement, connected to
the slips on the self-organized
critical behavior investigated in the last section. Each slip
reinitializes the values of $b$ and $d$ and the evolution towards the
critical attractor begins from scratch. It means that the long-range
correlations characteristic for the critical state cannot fully
develop. The difference can be seen already in the distribution of  
stresses, Fig. \ref{fig:distr}. The sharp edge in $P(b)$ observed in
the infinitely slow driving is smeared out. The position of the edge
determines the critical threshold $\lambda_c$ for the forward
avalanches, so we expect that no scaling of the type
(\ref{eq:aval-scaling}) will hold, as soon as the macroscopic movement has
non-zero velocity. 

\begin{figure}[hb]
  \centering
  \vspace*{90mm}
  \includegraphics{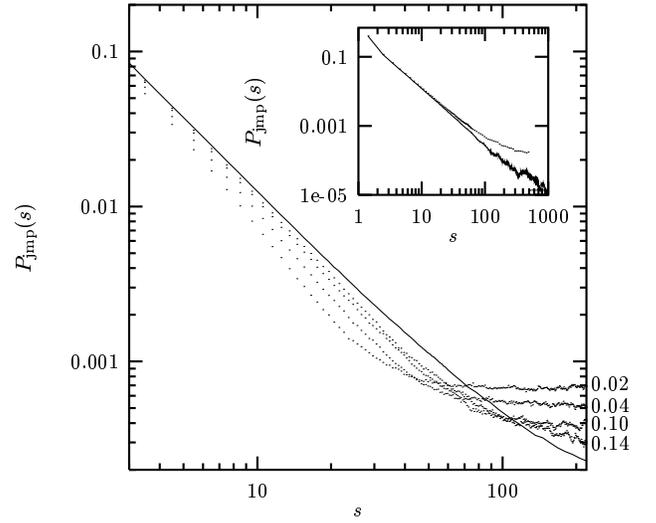}
  \caption{Distribution of jump lengths, for $N=10^3$, $\delta =
 0.001$, $b_M=0.9$. The full
 line is case without slips ($E_{\rm thr} = +\infty$). Dotted lines
 have slips allowed and the values of $E_{\rm thr}/N$ are indicated
 next to the position where the lines reach the right edge of the
 figure. In the inset, distribution of jump lengths, for $\delta=0$ and
 $N=10^4$ (full line) and $N=10^3$ (dotted line). Note that the inset
 makes it clear that the upward bent in the distribution for $E_{\rm
 thr} = +\infty$ is mere finite-size effect.}
  \label{fig:jumps}
\end{figure}

However, the most direct way how to investigate the
breakdown of criticality due to the slips seems to us to be the
calculation of the distribution of jump lengths. If in certain time
step $t$ the maximum stress was found at site $i_t$ and in the next
step at site $i_{t+1}$, we can compute spatial distance between this
sites as follows. Let $r_t(i)$ be the function which determines the
connectivity in time $t$, namely $r_t(i)$ is the site connected to $i$
on the right-hand side. The jump length $s$ is defined as
follows: starting from $i_t$ and applying $r_t$ we come to the right
neighbor of the extremal site at time $t$, $r_t(i_t)$. Then, applying
$s-l$ times 
the function $r_{t+1}$ we must end at $i_{t+1}$. So, $s$ is such that 
$i_{t+1}=(r_{t+1}^{~~\circ (s-1)}\circ r_t)(i_t)$.

In the self-organized critical state the probability distribution of
jump lengths is power law, $P_{\rm jmp}(s)=s^{-\pi}$. For $E_{\rm
thr}=\infty$ it is actually observed in our model, as indicated in the
inset in the Fig. \ref{fig:jumps}. The comparison of the
distributions for $N=10^3$ and $N=10^4$ is shown in order to give idea
of the magnitude of the finite-size corrections to the power-law
behavior.

The situation with non-zero macroscopic velocity, $E_{\rm thr} <
E_\infty$, is shown in the main Fig. \ref{fig:jumps}. When  $E_{\rm
thr}$ decreases, the velocity increases and the scale on which $P_{\rm
jmp}(s)$ obeys a power law shrinks. The correlations do not have time
enough to develop on the scale of the whole system, but only on
shorter distances. So, we may connect the velocity dependence of the
friction force to the level of correlations between the asperities,
which are present in the system. Contrary to the theories where the
velocity dependence stems from the aging of a single asperity, here the
aging is a collective effect. The age corresponds to the range of
correlations. For zero velocity the correlation length is infinite and
the age is infinite as well.

\section{Conclusions}

We presented a model of dry friction based on the conception of slider
and track interacting through a system of asperities. We proposed an
extremal dynamics model in order to describe the processes during the
movement of the slider. We found the decrease of the friction force
with increasing velocity. For velocities approaching to zero, the
friction force has finite limit. The origin of the velocity
dependence is not in a change of properties of a single asperity, but
in collective effects, involving many asperities. At zero velocity, the
system is in a highly correlated, self-organized critical state. The
value of the exponents are close to the Sneppen interface model,
however it is not clear from our data, whether the universality class
is the same.

Increasing velocity gradually destroys
the correlations. It is possible to view the buildup of the
correlations as a collective asperity aging mechanism, as a
counterpart to the single asperity aging due to plastic deformation. 
Such collective aging leads to different velocity dependence of the
friction force than in the models considernig single asperity aging
and may be thus tested experimentally. Two-dimensional variant of our
1D model would be
necessary for real comparison. However, the generalization to
arbitrary dimension is straightforward.

This observation reveals also the limits of applicability of
our model. It is appropriate in situations where the plastic
deformation 
does not dominate. 
The model is applicable to the regime of very small
velocities, where the usual logarithmic velocity dependence is
inappropriate. It may be also used to describe friction over highly
elastic surfaces, like rubber or some plastics, where the slow aging
of single asperities may not be dominant.

\vspace{5mm}
{\large\bf Acknowledgments}
I wish to thank M. Kotrla and  B. Velick\'y for useful
discussions.

\end{document}